\documentclass[conference,a4paper]{IEEEtran}

\usepackage{latexsym}
\usepackage{graphicx}
\usepackage{amsfonts,amssymb,amsmath}
\usepackage{ctable} 
\usepackage{mathtools, cuted}
\usepackage{hyperref}
\usepackage[ruled,linesnumbered,noend]{algorithm2e}
\usepackage[T1]{fontenc}
\usepackage{cite}
\usepackage{subcaption}
\usepackage{comment}
\usepackage{diagbox}

\usepackage{amsthm}
\usepackage{overpic}
\usepackage{steinmetz}
\usepackage{array}
\usepackage{url}
\usepackage{color}

\usepackage{algorithmicx}
\usepackage{algcompatible}

\usepackage{xcolor}
\usepackage{soul}

\usepackage{linegoal}

\theoremstyle{plain}

\newtheorem{remark}{Remark}

\newcommand{\vect}[1]{\mathbf{#1}}
\newcommand{\maximize}[1]{{\underset{{#1}}{\mathrm{maximize}}}}
\newcommand{\minimize}[1]{{\underset{{#1}}{\mathrm{minimize}}}}

\newcommand*{\LongState}[1]{\STATE
\parbox[t]{0.9\linewidth-\algorithmicindent-\algorithmicindent}{#1\strut}}

\def\Htran{\mbox{\tiny $\mathrm{H}$}}
\def\Ttran{\mbox{\tiny $\mathrm{T}$}}
\def\CN{\mathcal{C}\mathcal{N}} 

\SetCommentSty{mycommfont}

\IEEEoverridecommandlockouts

\begin{document}
\bstctlcite{BSTcontrol}
\title{Cell-Free MIMO in Space: Cooperative Satellite Transmission with Multi-Antenna Ground Users}

\author{Parisa Ramezani and Emil Bj\"{o}rnson\\
\IEEEauthorblockA{\textit{ Department of Computer Science, KTH Royal Institute of Technology, Stockholm, Sweden} \\  Email: \{parram, emilbjo\}@kth.se}%
\thanks{This work is funded by the SSF research center SMART-6GSAT (Sustainable Mobile Autonomous and Resilient 6G SatCom), reg.nr. CSG23-0001.}
\vspace{-3mm}}

\maketitle
\begin{abstract}
This paper develops a multi-user downlink communication framework for distributed low Earth orbit satellite networks serving ground users equipped with multiple antennas. Building upon the concept of cell-free multiple-input multiple-output in terrestrial networks, we propose a coordinated transmission scheme where multiple satellites jointly transmit spatially multiplexed data streams to each user. 
Using a new approximate achievable rate expression, we formulate a sum rate maximization problem under per-satellite and per-antenna power constraints and use the classical equivalence between sum rate maximization and mean square error minimization to optimize the satellites' precoding matrices using statistical channel state information. We numerically examine the performance of the proposed scheme in different settings and validate its effectiveness by comparing it against traditional precoding designs.
\end{abstract}


\section{Introduction}
Ubiquitous connectivity has been recognized by the International Telecommunication Union (ITU) as a key use case for 6G networks, aiming to close the digital divide \cite{wp5d2023m}. To achieve this, innovative approaches are needed to provide seamless service to underserved areas where traditional terrestrial infrastructure is not economically or logistically viable.
In this context, satellite communication is anticipated to play a pivotal role \cite{perez2019signal}. Among the various types of satellites, low Earth orbit (LEO) satellites have attracted increasing attention due to their unique advantages over geostationary and medium Earth orbit satellites, including significantly lower latency, reduced signal degradation, and lower production and launch costs. 

The application of multiple-input multiple-output (MIMO) transmission to satellite communications has recently attracted interest as a means to enhance data rates through advanced adaptive beamforming.
In \cite{you2020massive}, a massive MIMO framework for LEO satellites based on statistical channel state information (sCSI) is proposed to overcome the difficulty of acquiring instantaneous CSI in fast-moving LEO scenarios. Expanding on this, \cite{you2022hybrid} investigates hybrid analog/digital precoding to improve energy efficiency. The authors of \cite{alsenwi2024robust} propose a two-stage deep reinforcement learning approach for robust beamforming under outdated CSI caused by satellite mobility. 

The concept of cell-free MIMO, long studied in terrestrial networks \cite{demir2021foundations}, is now gaining traction in satellite communications to enable coordinated transmission from distributed satellites to ground users. \cite{humadi2024distributed} studies a user-centric distributed massive MIMO for LEO satellite networks, and proposes dynamic satellite clustering and phase-aware precoding. 
 The authors in \cite{ha2024user} focus on determining the most effective subset of beams for communication, while minimizing total transmission power and satisfying signal-to-interference-plus-noise ratio (SINR) constraints.
\cite{wang2025multiple} presents a dual-functional LEO satellite constellation framework that supports both information communication and location sensing through cooperative transmission from multiple satellites. 
 A distributed beamforming architecture for networked LEO satellites is presented in \cite{Zhang2025Enabling}, which optimizes beamformers based on sCSI.

 These studies have shown the promising potential of coordinated transmission from multiple satellites to ground users. Yet, they all assume single-antenna ground users, limiting spatial multiplexing capabilities. As modern user devices such as smartphones are equipped with multiple antennas, it is important to study satellite communication systems with multi-antenna ground users. 
 This paper investigates a distributed downlink satellite communication system, where multiple satellites jointly serve multi-antenna ground users by transmitting multiple spatial data streams to each user. The objective is to maximize the sum rate via precoding optimization at the satellites.  
Using a new tractable achievable rate expression, we optimize the precoding matrices under per-satellite transmit power constraints based on sCSI. 
Moreover, we extend the precoder design to enforce per-antenna power constraints, ensuring hardware-compliant precoders for each antenna port.
Numerical simulations show that joint transmission from geographically separated satellites enables per-user multi-stream, unlike a single co-located array with the same antenna count.
We further evaluate the performance of the proposed scheme for both per-satellite and per-antenna power constraints across various settings, and demonstrate the effectiveness of the proposed precoding scheme over traditional precoding designs. 

\vspace{-1mm}
\section{System Model}
A downlink multi-user distributed satellite communication (MU-D-SATCOM) system is considered where $L$ distributed LEO satellites (SATs), each equipped with $N$ antennas, serve $K$ user equipments (UEs), each having $M$ antennas. Since the UEs have multiple antennas, multiple spatially multiplexed data streams can be sent to each UE, if the channel supports it. Let $\vect{x}_k \sim \CN (\vect{0},\vect{I}_M) \in \mathbb{C}^{M}$ denote the data streams for UE\,$k$, where the signals for different UEs are independent
from each other and from the receiver noises. With centralized joint transmission, the received signal at UE\,$k$ is
\begin{align}
\label{eq:received_signal_1}
  \vect{y}_k = \sum_{l=1}^L \vect{H}_{l,k} \vect{W}_{l,k} \vect{x}_k + \sum_{i = 1, i \neq k}^K \sum_{l=1}^L \vect{H}_{l,k}\vect{W}_{l,i}\vect{x}_i + \vect{n}_k,   
\end{align}
where $\vect{H}_{l,k} \in \mathbb{C}^{M \times N}$ is the channel between SAT\,$l$ and UE\,$k$, $\vect{W}_{l,k} \in \mathbb{C}^{N \times M }$ is the precoding matrix of SAT\,$l$ for UE\,$k$'s data, and $\vect{n}_k \sim \CN(0, \sigma^2\vect{I}_M)$ is the noise vector at UE\,$k$.

\subsection{Channel Model}
We consider the typical scenario in satellite communication: far-field line-of-sight (LoS) channels between the SATs and UEs. We further assume that the antenna panels at the SATs and UEs are in the form of uniform linear arrays (ULAs). Under these assumptions, the channel $\vect{H}_{l,k}$ can be modeled as 
\begin{align}
\label{eq:geometric_channel}
  \vect{H}_{l,k} = \gamma_{l,k} \vect{b}_k(\theta_{l,k})\vect{a}_l^{\Ttran}(\varphi_{l,k}),  
\end{align}
where $\gamma_{l,k}$ is the complex channel gain between SAT\,$l$ and UE\,$k$, $\vect{b}_k(\cdot)$ and $\vect{a}_l(\cdot)$ are respectively the array response vectors at UE\,$k$ and SAT\,$l$, and $\theta_{l,k}$ and $\varphi_{l,k}$ are the respective angle-of-arrival (AoA) and angle-of-departure (AoD).  The SATs typically have precise orbital data and the UEs' locations can be reliably tracked via GPS. Thus, it is reasonable to assume that all AoAs' and AoDs' information is available \cite{you2020massive,you2022hybrid,Zhang2025Enabling}. However,  channel gain $\gamma_{l,k}$
 varies rapidly; thus, only its statistical information is usually available. 
 We model $\gamma_{l,k}$ as Rician fading with a time-varying random LoS phase and variance $\beta_{l,k}$. Specifically, 
 \begin{equation}
  \gamma_{l,k} = \sqrt{\beta_{l,k}} \left(\sqrt{\frac{\kappa_{l,k}}{\kappa_{l,k}+1}} e^{j\psi_{l,k}} + \sqrt{\frac{1}{\kappa_{l,k} + 1}} z_{l,k} \right),   
 \end{equation}
 where $\psi_{l,k} \sim \mathcal{U}[0,2\pi)$, $z_{l,k} \sim \CN(0,1)$, $\psi_{l,k}$ and $z_{l,k}$ are independent for all $l,k$, and $\beta_{l,k}$ is known a priori. 
\vspace{-1mm}
\subsection{ Problem Formulation}
Based on the received signal expression in \eqref{eq:received_signal_1}, the achievable rate at UE\,$k$ is given by \eqref{eq:achievable_rate}, shown at the top of the next page, where the expectation is over different channel realizations.   
\begin{figure*}[t]
    \begin{align}
    \label{eq:achievable_rate}
  R_k = \mathbb{E}\left\{\log_2 \left| \vect{I}_M + \left(\sum_{l=1}^L \vect{H}_{l,k} \vect{W}_{l,k}\right) \left(\sum_{l=1}^L \vect{H}_{l,k} \vect{W}_{l,k}\right)^{\Htran} \left(\sum_{i=1,i\neq k}^K \left(\sum_{l=1}^L \vect{H}_{l,k}\vect{W}_{l,i} \right)\left(\sum_{l=1}^L \vect{H}_{l,k}\vect{W}_{l,i} \right)^{\Htran}   + \sigma^2\vect{I}_M\right)^{-1}\right| \right\}  
\end{align}
    \hrulefill
\end{figure*}
The objective is to maximize the sum rate of all UEs by optimizing the precoding matrices at each SAT, subject to individual total power constraints at the SATs. The problem of interest is thus formulated as 
\vspace{-1.5mm}
\begin{subequations}
\label{eq:main_problem}
 \begin{align}
  &\maximize{\{\vect{W}_{l,k}\}_{\forall l,k}}\,\, \sum_{k=1}^K R_k, \\
    &\mathrm{subject~to}\, \sum_{k=1}^K \mathrm{Tr}\left(\vect{W}_{l,k}^{\Htran}\vect{W}_{l,k} \right) \leq \rho_l,~\forall l, \label{eq:power_constraint}
    \end{align}
 \end{subequations}
where $\rho_l$ is the maximum transmit power for SAT\,$l$. Note that as $R_k$ in \eqref{eq:achievable_rate} is an expectation over the fading, the optimized precoders are independent of $\gamma_{l,k}$, and only depend on the angular information. $R_k$ does not have an analytical form, which makes problem \eqref{eq:main_problem} difficult to handle. In the following, we will use an approximate form of \eqref{eq:achievable_rate} to reformulate problem \eqref{eq:main_problem}.

\section{Problem Reformulation and Precoding Design}.  
As the first step to solve \eqref{eq:main_problem}, we seek to reformulate the objective function in a more tractable form. To this end, we adopt an approximation to simplify the achievable rate. For positive definite matrices $\vect{X}$ and $\vect{Y}$, with $|\cdot|$ denoting the determinant, it holds that 
\begin{align}
\label{eq:approximation}
    \mathbb{E}\left\{\log_2 \left| \vect{I} + \vect{X}\vect{Y}^{-1}  \right|\right\}  \approx \log_2 \left|\vect{I} + \mathbb{E}\{\vect{X}\} \mathbb{E}\{\vect{Y}\}^{-1}  \right|.
  \end{align}

  \begin{remark}
  The approximation in \eqref{eq:approximation} has been shown to be tight for scalar cases \cite{zhang2014power}. However, to the best of our knowledge, there is no mathematical proof establishing its tightness in the matrix-valued case. Our numerical evaluations indicated that the approximation tightness can vary; nevertheless, the right-hand side (RHS) provides a reasonable approximation to the left-hand side (LHS) as both exhibit consistent trends. 
 Section~\ref{sec:results} offers a detailed numerical comparison of the LHS and RHS of \eqref{eq:approximation}.
\end{remark}

\vspace{-3mm}
\subsection{Problem Reformulation} 
 The achievable rate expression can be simplified using \eqref{eq:approximation}. In particular, for the signal term we have 
 \begin{align}
 \label{eq:signal_approx}
  &\mathbb{E}\left\{\left(\sum_{l=1}^L \vect{H}_{l,k} \vect{W}_{l,k}\right) \left(\sum_{l=1}^L \vect{H}_{l,k} \vect{W}_{l,k}\right)^{\Htran}\right\}  \nonumber \\
  &  = \mathbb{E}\left\{\sum_{l=1}^L |\gamma_{l,k}|^2 \vect{b}_{l,k} \vect{a}_{l,k}^{\Ttran} \vect{W}_{l,k} \vect{W}_{l,k}^{\Htran} \vect{a}_{l,k}^* \vect{b}_{l,k}^{\Htran}\right\} \nonumber \\
  &  = \sum_{l=1}^L \mathbb{E}\left\{|\gamma_{l,k}|^2\right\} \vect{b}_{l,k}\vect{a}_{l,k}^{\Ttran} \vect{W}_{l,k} \vect{W}_{l,k}^{\Htran} \vect{a}_{l,k}^* \vect{b}_{l,k}^{\Htran} \nonumber \\
  & = \sum_{l=1}^L \beta_{l,k} \vect{b}_{l,k}\vect{a}_{l,k}^{\Ttran} \vect{W}_{l,k} \vect{W}_{l,k}^{\Htran} \vect{a}_{l,k}^* \vect{b}_{l,k}^{\Htran},
 \end{align}
where we use the notation $\vect{b}_{l,k} = \vect{b}_k(\theta_{l,k})$ and $\vect{a}_{l,k} = \vect{a}_k(\varphi_{l,k})$. For the interference-plus-noise term, we obtain
\begin{align}
\label{eq:interference_approx}
 &\mathbb{E}\left\{\sum_{i=1,i\neq k}^K \left(\sum_{l=1}^L \vect{H}_{l,k}\vect{W}_{l,i} \right)\left(\sum_{l=1}^L \vect{H}_{l,k}\vect{W}_{l,i} \right)^{\Htran}   + \sigma^2\vect{I}_M  \right\}  \nonumber \\ 
 & = \sum_{i=1,i\neq k}^K \sum_{l=1}^L \beta_{l,k}\vect{b}_{l,k}\vect{a}_{l,k}^{\Ttran} \vect{W}_{l,i} \vect{W}_{l,i}^{\Htran} \vect{a}_{l,k}^* \vect{b}_{l,k}^{\Htran} + \sigma^2 \vect{I}_M.
\end{align}
Therefore, the achievable rate expression $R_k$ in \eqref{eq:achievable_rate} can be approximated as $\bar{R}_k$ in \eqref{eq:achievable_rate_approx}, shown at the top of the next page.%
\begin{figure*}[!t]
    \begin{align}
    \label{eq:achievable_rate_approx}
  \bar{R}_k = \log_2 \left| \vect{I}_M + \left( \sum_{l=1}^L \beta_{l,k} \vect{b}_{l,k}\vect{a}_{l,k}^{\Ttran} \vect{W}_{l,k} \vect{W}_{l,k}^{\Htran} \vect{a}_{l,k}^* \vect{b}_{l,k}^{\Htran} \right) \left( \sum_{i=1,i\neq k}^K \sum_{l=1}^L \beta_{l,k}\vect{b}_{l,k}\vect{a}_{l,k}^{\Ttran} \vect{W}_{l,i} \vect{W}_{l,i}^{\Htran} \vect{a}_{l,k}^* \vect{b}_{l,k}^{\Htran} + \sigma^2 \vect{I}_M \right)^{-1} \right| 
\end{align}
    \hrulefill
\end{figure*}
The sum rate maximization problem is thus reformulated as 
\vspace{-2mm}
\begin{subequations}
\label{eq:approximate_problem}
 \begin{align}
  &\maximize{\{\vect{W}_{l,k}\}_{\forall l,k}}\,\, \sum_{k=1}^K \bar{R}_k, \\
    &\mathrm{subject~to}\, \sum_{k=1}^K \mathrm{Tr}\left(\vect{W}_{l,k}^{\Htran}\vect{W}_{l,k} \right) \leq \rho_l,~\forall l. 
    \end{align}
 \end{subequations}
 \subsection{Precoding Design}
 Problem \eqref{eq:approximate_problem} is still non-convex due to the non-concavity of the objective function and the coupling between the precoding matrices. 
 Weighted minimum mean square error (WMMSE) is a well-established method for solving non-convex sum-rate maximization problems involving interference terms \cite{Shi2011}. We thus resort to this method to solve \eqref{eq:approximate_problem}. 
 Since \eqref{eq:approximate_problem} is based on the approximate rate in \eqref{eq:achievable_rate_approx}, which depends on the channels only through the effective matrices $\tilde{\vect{H}}_{l,k}=\sqrt{\beta_{l,k}}\vect{b}_{l,k}\vect{a}_{l,k}^{\Ttran}$, we perform the WMMSE derivations using the corresponding effective received signal obtained from \eqref{eq:received_signal_1} by replacing $\vect{H}_{l,k}$ with $\tilde{\vect{H}}_{l,k}$, i.e., $
\tilde{\vect{y}}_k=\sum_{l=1}^{L}\tilde{\vect{H}}_{l,k}\sum_{i=1}^{K}\vect{W}_{l,i}\vect{x}_i+\vect{n}_k$.
 Assuming that UE\,$k$ uses the receive combining matrix $\vect{U}_k \in \mathbb{C}^{M \times M}$ to decode its desired signal, the estimated signal at UE\,$k$ is given by $\hat{\vect{x}}_k = \vect{U}_k^{\Htran}\Tilde{\vect{y}}_k$.   The MSE matrix for UE\,$k$ is
 \begin{align}
 \label{eq:MSE}
 &\vect{E}_k = \mathbb{E}\left\{ (\hat{\vect{x}}_k - \vect{x}_k)(\hat{\vect{x}}_k - \vect{x}_k)^{\Htran} \right\} = \nonumber \\
 & \vect{U}_k^{\Htran} \left (\sum_{i=1}^K \sum_{l=1}^L \Tilde{\vect{H}}_{l,k} \vect{W}_{l,i} \vect{W}_{l,i}^{\Htran} \Tilde{\vect{H}}_{l,k}^{\Htran}\right) \vect{U}_k \nonumber \\ &-2\Re\left(\vect{U}_k^{\Htran} \sum_{l=1}^L \Tilde{\vect{H}}_{l,k}\vect{W}_{l,k}\right) + \sigma^2 \vect{U}_k^{\Htran} \vect{U}_k + \vect{I}_M,
 \end{align}
 
 The sum rate maximization problem in \eqref{eq:approximate_problem} can now be formulated as an equivalent weighted MSE minimization problem following an approach from \cite{Shi2011}. Specifically, letting $\vect{C}_k \in \mathbb{C}^{M \times M}$ be a weight matrix associated with $\vect{E}_k$, we can solve the following problem:  
 \begin{subequations}
\label{eq:WMMSE_problem}
 \begin{align}
  \minimize{\{\vect{U}_k\}_{\forall k}, \{\vect{C}_k\}_{\forall k}, \{\vect{W}_{l,k}\}_{\forall l,k}}\,\, &\sum_{k=1}^K \mathrm{Tr}\left(\vect{C}_k \vect{E}_k \right) - \log_2 \left|\vect{C}_k \right|, \label{eq:WMMSE_OF}\\
    ~~~~~~~~~\mathrm{subject~to}\, ~~~&\sum_{k=1}^K \mathrm{Tr}\left(\vect{W}_{l,k}\vect{W}_{l,k}^{\Htran} \right) \leq \rho_l,~\forall l. 
    \end{align}
 \end{subequations}
 Problem \eqref{eq:WMMSE_problem} is equivalent to problem \eqref{eq:approximate_problem} in that the optimal precoding matrices $\left\{\vect{W}_{l,k}\right\}$ are the same. Yet, \eqref{eq:WMMSE_problem} is more tractable than \eqref{eq:approximate_problem} as it is convex with respect to each block of variables, i.e., $\{\vect{U}_k\}$, $\{\vect{C}_k\}$, and $\{\vect{W}_{l,k}\}$. This motivates the use of the block coordinate descent method since the per-block convexity ensures convergence to a stationary point \cite{Tseng2001}.  

 Assuming fixed $\{\vect{W}_{l,k}\}$, the optimal receive beamforming matrix can be readily obtained by setting $\frac{\partial \mathrm{Tr}( \vect{E}_k)}{\partial \vect{U}_k} = \vect{0}$ for each individual UE. This results in
 \vspace{-2mm}
 \begin{align}
 \label{eq:receive_beamforming}
  \vect{U}^\star_k = \left(\sum_{i=1}^K \Tilde{\vect{H}}_k \vect{W}_i \vect{W}_i^{\Htran} \Tilde{\vect{H}}_k^{\Htran} + \sigma^2 \vect{I}_M \right)^{-1} \Tilde{\vect{H}}_k \vect{W}_i,
 \end{align}
 where $\Tilde{\vect{H}}_k = [\Tilde{\vect{H}}_{1,k},\ldots,\Tilde{\vect{H}}_{L,k}] \in \mathbb{C}^{M \times LN}$ and $\vect{W}_i = \left[\vect{W}_{1,i}^{\Ttran},\ldots,\vect{W}_{L,i}^{\Ttran}\right]^{\Ttran} \in \mathbb{C}^{LN \times M}$. 
Then, assuming $\{\vect{U}_k\}$ and $\{\vect{W}_{l,k}\}$ are fixed, the optimal weight matrix $\vect{C}_k$ can be found by solving $\frac{\partial \left( \mathrm{Tr}\left(\vect{C}_k \vect{E}_k - \log_2 |\vect{C}_k|\right)\right)}{\partial \vect{C}_k} = \vect{0}$ for $\vect{C}_k$, resulting in
\vspace{-1.5mm}
\begin{align}
\label{eq:weight_matrix}
  \vect{C}_k^\star = \frac{1}{\mathrm{ln}\,2}\, \vect{E}_k^{-1},  
\end{align}
where $\vect{E}_k$ is obtained from \eqref{eq:MSE} after substituting $\vect{U}_k^\star$.
 The final step is to optimize $\{\vect{W}_{l,k}\}$, which can be decomposed into $L$ sub-problems with the $l$th sub-problem being
 \vspace{-1mm}
 \begin{subequations}
\label{eq:precoding_optimization_separated}
 \begin{align}
  &\minimize{\{\vect{W}_{l,k}\}_{\forall l,k}}\,\, \sum_{k=1}^K \mathrm{Tr}\Bigg(\sum_{i=1}^K 
\vect{C}_i \vect{U}_i^{\Htran}   
\Tilde{\vect{H}}_{l,i} \vect{W}_{l,k} \vect{W}_{l,k}^{\Htran} 
\Tilde{\vect{H}}_{l,i}^{\Htran} \vect{U}_i   \nonumber \\
& - 2\vect{C}_k\Re\left(\vect{U}_k^{\Htran}  
\Tilde{\vect{H}}_{l,k}\vect{W}_{l,k}\right) \Bigg), \\
    &~\mathrm{subject~to}\, ~\sum_{k=1}^K \mathrm{Tr}\left(\vect{W}_{l,k}^{\Htran}\vect{W}_{l,k} \right) \leq \rho_l. \label{eq:total_power_constraint}
    \end{align}
 \end{subequations}
 We can now utilize the method of Lagrange multipliers to solve \eqref{eq:precoding_optimization_separated}. Specifically, the Lagrangian is given by 
 \begin{align}
   &\mathcal{L}_l = \sum_{k=1}^K \mathrm{Tr}\left( \vect{W}_{l,k}^{\Htran} \left(\sum_{i=1}^K \Tilde{\vect{H}}_{l,i}^{\Htran} \vect{U}_i \vect{C}_i \vect{U}_i^{\Htran} \Tilde{\vect{H}}_{l,i}\right) \vect{W}_{l,k} \right. -\nonumber \\
   &    2 \vect{C}_k \Re \left(\vect{U}_k^{\Htran}  
\Tilde{\vect{H}}_{l,k}\vect{W}_{l,k}\right) \Bigg) + \mu_l \left( \sum_{k=1}^K \mathrm{Tr}\left(\vect{W}_{l,k}^{\Htran} \vect{W}_{l,k}\right) - \rho_l \right),
 \end{align}
where $\mu_l \geq 0$ is the Lagrange multiplier for the power constraint at SAT\,$l$. We need to solve the following problem 
\vspace{-1.5mm}
\begin{align}
   \minimize{\{\vect{W}_{l,k}\}_{\forall k}}\, \mathcal{L}_l, 
\end{align}
which can be separately solved for the $K$ UEs by setting $\frac{\partial \mathcal{L}_l}{\partial \vect{W}_{l,k}} = \vect{0}$. The optimal $\vect{W}_{l,k}$ for given $\mu_l$ is obtained as 
\begin{align}
\label{eq:precoding_matrix}
 \vect{W}_{l,k}^\star (\mu_l) = \left(\sum_{i=1}^K \Tilde{\vect{H}}_{l,i}^{\Htran} \vect{U}_i \vect{C}_i \vect{U}_i^{\Htran}\Tilde{\vect{H}}_{l,i} + \mu_l \vect{I}_N\right)^{-1} \Tilde{\vect{H}}_{l,k}^{\Htran}\vect{U}_k \vect{C}_k.    
\end{align}

 To obtain the optimal value of $\mu_l$, we first examine $\mu_l = 0$. If $\sum_{k=1}^K \mathrm{Tr}(\vect{W}_{l,k}^{\star \Htran}(0) \vect{W}_{l,k}^{\star }(0) ) \leq \rho_l$, then $\mu_l^\star = 0$. Otherwise, a feasible upper bound is identified by geometric expansion: starting with any  $\mu_l$, the value is multiplied by $\alpha>1$ until the resulting total power satisfies the constraint. Then, bisection over $[0,\mu_l]$ is applied.

 The proposed precoding design is summarized in Algorithm~\ref{alg:WMMSE}, and the method for updating Lagrange multipliers is given in Algorithm~\ref{alg:bisection}. The precoding matrices in Algorithm~\ref{alg:WMMSE} are initialized as 
   $ \vect{W}_{l,k} = \sqrt{\rho_{l,k}} \frac{\Tilde{\vect{W}}_{l,k}}{\|\Tilde{\vect{W}}_{l,k}\|_{\mathrm{F}}}$, 
 where $\rho_{l,k} = \rho_l \frac{\sqrt{\beta_{l,k}}}{\sum_{i=1}^K \sqrt{\beta_{l,i}}}$, and 
   $\Tilde{\vect{W}}_{l,k} =  \left(\sum_{i=1}^K \Tilde{\vect{H}}_{l,i}^{\Htran} \Tilde{\vect{H}}_{l,i} + \sigma^2 \vect{I}_N\right)^{-1} \Tilde{\vect{H}}_{l,k}^{\Htran}$. 
 This initializer is the MMSE solution with a power scaling that slightly prioritizes UEs with better channels. 
 \begin{algorithm}
\small
\caption{\small Proposed precoding for MU-D-SATCOM }
\label{alg:WMMSE}
	\begin{algorithmic}[1]
 \STATEx{\textbf{Inputs:} $\varphi_{l,k}$, $\theta_{k,l}$, $\beta_{l,k}~\forall l,k$, $\sigma^2$, $\epsilon$, $I_{\mathrm{max}}$ .}
		
		\STATE {Initialize the precoding matrices $\vect{W}_{l,k}~\forall l,k$.}
		\STATE {Initialize $\delta = \epsilon + 1$ and set the initial value of the objective function in \eqref{eq:WMMSE_OF} to $+\infty$. }
	\STATE{Set $I = 0$.}	
        \WHILE{$\delta > \epsilon ~\&~ I < I_{\mathrm{max}}$ }

        \STATE{$I \leftarrow I+1$}.
        \STATE{Find receive beamforming matrices $\vect{U}_k^\star ~\forall k$ from \eqref{eq:receive_beamforming}.}
        \STATE{Find weight matrices  $\vect{C}^\star_k~\forall k$ from \eqref{eq:weight_matrix}. }
        \LongState{Find the precoding matrices from \eqref{eq:precoding_matrix} for $\mu_l^\star~\forall l$ obtained via Algorithm~\ref{alg:bisection}. }
        \LongState{Compute the objective function in \eqref{eq:WMMSE_OF} with the updated variables. }
        \LongState{Set $\delta$ as the difference between the current and previous objective function values. }
        \ENDWHILE
		\STATEx{\textbf{Output:} $\vect{W}_{l,k}^\star (\mu_l^\star)~\forall l,k$.}
	\end{algorithmic}
\end{algorithm} 

\begin{algorithm}[t]
\small
\caption{\small  Lagrange Multiplier Optimization }
\label{alg:bisection}
	\begin{algorithmic}[1]
 \STATEx{\textbf{Inputs:} $\Tilde{\vect{H}}_{l,k}$, $\vect{U}_k$, $\vect{C}_k~\forall k$, $\rho_l$, $\alpha$, $\varepsilon$.}
		
		\STATE {Set $\mu_l^{(L)} = 0$, $\mu_l^{(U)} = +\infty$, $\mathrm{flag} = 0$, and initialize $\mu_l$.}

        \WHILE{$(\mu_l^{(U)} - \mu_l^{(L)}) > \varepsilon $}
        \STATE{ Compute $\vect{W}_{l,k}^\star~\forall k$ from \eqref{eq:precoding_matrix}.}
		\IF{$\sum_{k=1}^K \mathrm{Tr}\left(\vect{W}_{l,k}^\star(\mu_l) \vect{W}_{l,k}^{\star \Htran}(\mu_l) \right) \leq \rho_l$}
         \STATE{$\mu_l^{(U)} \leftarrow \mu_{l}$, $\mu_l \leftarrow (\mu_l^{(L)} +  \mu_l^{(U)})/2 $.}
         \STATE{$\mathrm{flag} = 1$.}
         \ELSIF{$\sum_{k=1}^K \mathrm{Tr}\left(\vect{W}_{l,k}^\star(\mu_l) \vect{W}_{l,k}^{\star \Htran}(\mu_l) \right) > \rho_l~\& ~\mathrm{flag} = 1$ }
         \STATE{$\mu_{l}^{(L)} \leftarrow \mu_l$, $\mu_l \leftarrow (\mu_l^{(L)} +  \mu_l^{(U)})/2 $.}
         \ELSE
         \STATE{$\mu_l \leftarrow \alpha \mu_l$.}
        \ENDIF
		\ENDWHILE
        \STATEx{\textbf{Output:} $\mu_l^\star = \mu_l$.}
	\end{algorithmic}
  
\end{algorithm} 
  \vspace{-9mm}
\section{Precoding under Per-Antenna Constraints}
\vspace{-2mm}
In the previous section, we derived the optimal precoding matrices under a total power constraint at each SAT. However, in practice, the transmitters onboard each SAT are power-limited on a per-antenna-port 
basis, rather than by a single sum-power budget, because each antenna port is driven by its own power amplifier. The consideration of per-antenna power constraint is especially important for satellite communication, since once in orbit, the hardware cannot be repaired or replaced, and exceeding per-chain limits can cause irreversible damage. 
In this section, we revisit the precoding design by taking into account this practical constraint.  
To accommodate per-antenna power constraints, \eqref{eq:total_power_constraint} is modified as 
\begin{align}
 \sum_{k=1}^K \mathrm{Tr}\left(\vect{W}_{l,k}^{\Htran}\vect{E}_n \vect{W}_{l,k}\right) \leq \rho_{l,n},~~\forall l,n,  
\end{align}
where $\vect{E}_n \in \mathbb{C}^{N \times N}$ is a diagonal matrix with $1$ on its $n$th diagonal element and $0$ elsewhere. 

To incorporate the per-antenna power constraint into  precoding design, only the precoder optimization step needs to be modified, while other steps (i.e., obtaining receive beamforming and weight matrices) remain unchanged. In particular, after decomposing the precoding optimization problem into $L$ different sub-problems, we arrive at the new Lagrangian
 \begin{align}
   &\mathcal{L}_l = \sum_{k=1}^K \mathrm{Tr}\left( \vect{W}_{l,k}^{\Htran} \left(\sum_{i=1}^K \Tilde{\vect{H}}_{l,i}^{\Htran} \vect{U}_i \vect{C}_i \vect{U}_i^{\Htran} \Tilde{\vect{H}}_{l,i}\right) \vect{W}_{l,k} \right. -\nonumber \\
   &    2 \vect{C}_k \Re \left(\vect{U}_k^{\Htran}  
\Tilde{\vect{H}}_{l,k}\vect{W}_{l,k}\right) \Bigg) + \nonumber \\ &\sum_{n=1}^N \left(\mu_{l,n} \left( \sum_{k=1}^K \mathrm{Tr}\left(\vect{W}_{l,k}^{\Htran} \vect{E}_n \vect{W}_{l,k}\right) - \rho_{l,n} \right) \right),
 \end{align}
where $\mu_{l,n} \geq 0$ is the Lagrange multiplier associated with the power constraint at the $n$th antenna of SAT\,$l$. Similar to the previous section, minimizing the Lagrangian can be separately done for the $K$ UEs by setting $\frac{\partial \mathcal{L}_l}{\partial \vect{W}_{l,k}} = \vect{0}$, resulting in 
\begin{align}
\label{eq:precoding_matrix2}
 &\vect{W}_{l,k}^\star (\boldsymbol{\mu}_l) = \nonumber \\ &\left(\sum_{i=1}^K \Tilde{\vect{H}}_{l,i}^{\Htran} \vect{U}_i \vect{C}_i \vect{U}_i^{\Htran}\Tilde{\vect{H}}_{l,i} + \sum_{n=1}^N\mu_{l,n} \vect{E}_n\right)^{-1} \Tilde{\vect{H}}_{l,k}^{\Htran}\vect{U}_k \vect{C}_k,    
\end{align}
where $\boldsymbol{\mu}_l = [\mu_{l,1},\ldots,\mu_{l,N}]^{\Ttran}$ can be updated using the ellipsoid method which extends the bisection method to multiple dimensions by iteratively refining an ellipsoidal region containing the optimal solution \cite{grotschel2012geometric}. Furthermore, the $n$th row of the initial precoder $\vect{W}_{l,k}$  will be $\vect{W}_{l,k}[n,:] = \sqrt{\rho_{l,n,k}} \frac{\Tilde{\vect{W}}_{l,k}[n,:]}{\left\|\Tilde{\vect{W}}_{l,k}[n,:]\right\|}$,  
where $\rho_{l,n,k} = \rho_{l,n} \frac{\sqrt{\beta_{l,k}}}{\sum_{i=1}^K \sqrt{\beta_{l,i}}}$ and  $\Tilde{\vect{W}}_{l,k}$ has been defined preceding Algorithm~\ref{alg:WMMSE}.

\section{Numerical Results}
\label{sec:results}
\begin{figure}[t!]
  \centering
   \begin{overpic}[width=0.65\linewidth]{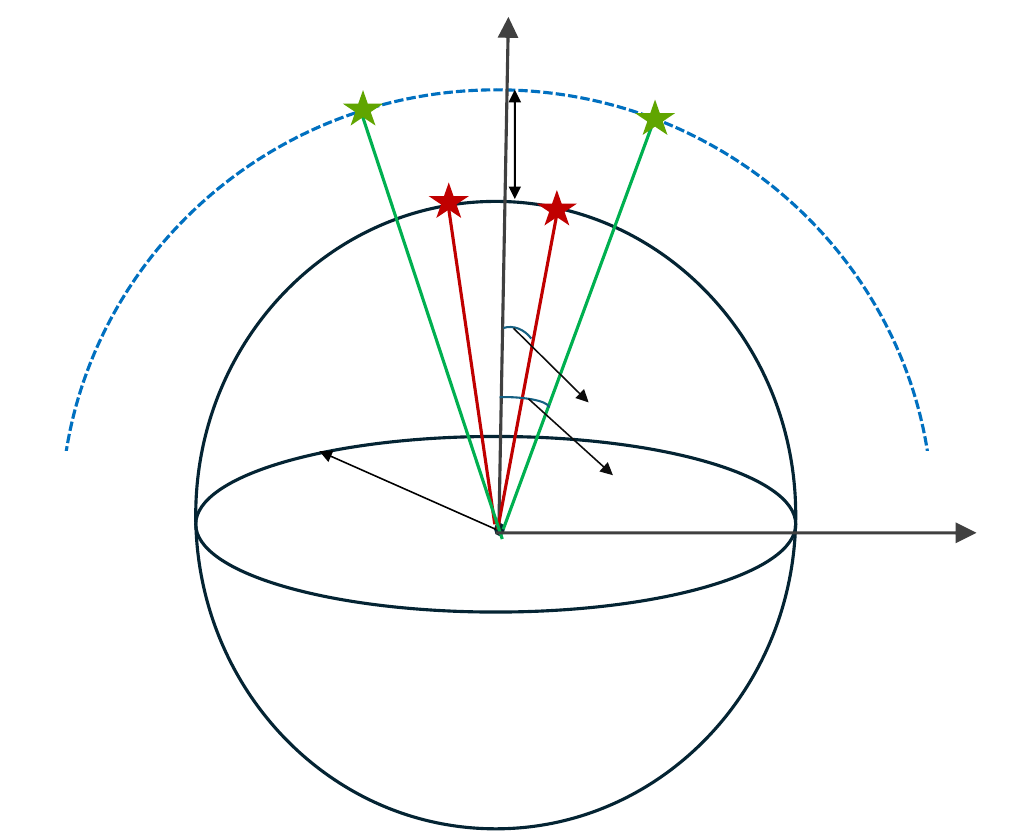}
  \put(32,31.5){\small $R_{\mathrm{E}}$}%
  \put(52.4,67){$h$}%
  \put(61,33){$\vartheta_{\mathrm{s}}$}%
  \put(57.8,39.5){$\vartheta_{\mathrm{u}}$}%
  \put(92,25){ \large{$x$}}%
  \put(51.5,77.7){ \large{$z$}}
   \end{overpic}
\caption{Simulation layout of the proposed MU-D-SATCOM.}
\label{fig:setup}
\vspace{-4mm}
\end{figure}
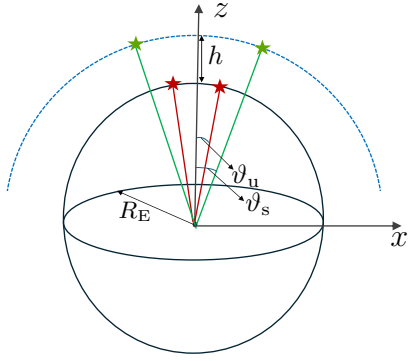

In this section, we evaluate the performance of the presented schemes. For brevity and visualization clarity, we consider a 2D simulation setup, as shown in Fig.\,~\ref{fig:setup}. 
We consider an Earth-centered Earth-fixed (ECEF) coordinate system with the Earth center as the origin and the Earth radius being $R_{\mathrm{E}}$. 
The UEs are randomly placed within an angular range $[-\vartheta_{\mathrm{u}},\vartheta_{\mathrm{u}}]$ relative to the vertical line extending from the Earth center (i.e., the zenith direction), as illustrated by red stars. 
The SATs are positioned at a fixed orbital altitude $h$ and distributed with equal angular spacing within the range $[-\vartheta_{\mathrm{s}},\vartheta_{\mathrm{s}}]$ around the zenith direction, as shown by green stars. 
Assuming that the angles of UE\,$k$ and SAT\,$l$ with respect to the zenith direction are  $\vartheta_{\mathrm{u},k}$ and $\vartheta_{\mathrm{s},l}$, respectively, their positions in the $xz$ plane can be obtained as $\vect{p}_{\mathrm{u},k} = [R_{\mathrm{E}}\sin(\vartheta_{\mathrm{u},k}),R_{\mathrm{E}}\cos(\vartheta_{\mathrm{u},k})]^{\Ttran}$ and $\vect{p}_{\mathrm{s},l} = [(R_{\mathrm{E}}+h)\sin(\vartheta_{\mathrm{s},l}),(R_{\mathrm{E}}+h)\cos(\vartheta_{\mathrm{s},l})]^{\Ttran}$, respectively. The corresponding AoD and AoA are then computed as 
\begin{align}
  \varphi_{l,k} = \tan^{-1}\left(\frac{\vect{s}_{l,\perp}^{\Ttran}\vect{v}_{l\rightarrow k}}{\vect{s}_l^{\Ttran
  }\vect{v}_{l\rightarrow k}}\right),
\theta_{l,k} = \tan^{-1}\left(\frac{\vect{u}_{k,\perp}^{\Ttran}\vect{v}_{k\rightarrow l}}{\vect{u}_k^{\Ttran
  }\vect{v}_{k\rightarrow l}}\right),
\end{align}
where $\vect{s}_l = -\vect{p}_{\mathrm{s},l}/\|\vect{p}_{\mathrm{s},l}\| = [s_{l,x},s_{l,z}]^{\Ttran}$ is the boresight unit vector for SAT\,$l$ and  $\vect{s}_{l,\perp} = [-s_{l,z},s_{l,x}]^{\Ttran} $ is the vector perpendicular to it.  Similarly, $\vect{u}_k = \vect{p}_{\mathrm{u},k}/\|\vect{p}_{\mathrm{u},k}\| = [u_{k,x},u_{l,z}]^{\Ttran}$  and  $\vect{u}_{k,\perp} = [-u_{k,z},u_{k,x}]^{\Ttran} $. Furthermore, $\vect{v}_{l \rightarrow k} = \vect{p}_{\mathrm{u},k} - \vect{p}_{\mathrm{s},l}$ and $\vect{v}_{k \rightarrow l} = \vect{p}_{\mathrm{s},l} - \vect{p}_{\mathrm{u},k}$
are the direction vectors. 

 Unless otherwise stated, these parameters are used for the simulations: $N = 8$ antennas per SAT, $K = 8$ UEs, $M = 2$ antennas per UE, $\rho_l= \rho = 50$\,W ($\approx 47$\,dBm) $\forall l$, $\sigma^2 = -124\,$dBm, and $\vartheta_{\mathrm{u}} = 1^\circ$. The Earth radius is $R_{\mathrm{E}} = 6371$\,km and the SATs' orbit altitude is $h = 500$\,km. The large-scale fading coefficients are
\begin{align}
  \beta_{l,k} = G_{\mathrm{s}} G_{\mathrm{u}} \left(\frac{\nu_c}{4 \pi f_c d_{l,k}}\right)^2, \forall l,k, 
\end{align}
where $G_{\mathrm{s}} = 6\,$dBi and $G_{\mathrm{u}} = 0\,$dBi are the antenna gains of the SATs and the UEs, $\nu_c = 3\cdot 10^8$ is the speed of light, and $f_c = 8\,$GHz is the carrier frequency.  Furthermore, $d_{l,k} = \|\vect{p}_{\mathrm{s},l} - \vect{p}_{\mathrm{u},k}\|$ is the distance between SAT\,$l$ and UE\,$k$. The Rician factor is set as $\kappa_{l,k} = 12\,$dB\,$\forall l,k$. Other parameters in Algorithms~\ref{alg:WMMSE} and \ref{alg:bisection} are set as $\epsilon = 10^{-4}$, $I_{\mathrm{max}} = 200$, $\varepsilon = 10^{-3}$, and $\alpha = 2$. 

\begin{figure}
	\centering
	\begin{subfigure}{0.67\linewidth}
		\centering
		\includegraphics[width=\columnwidth]{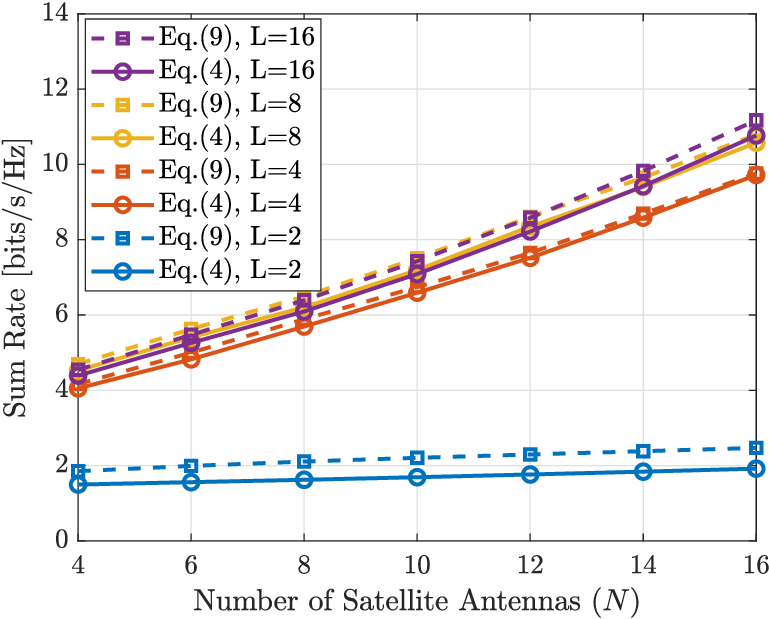}
		\caption{$M=2$.}
		
	\end{subfigure}
	\begin{subfigure}{0.67\linewidth}
		\centering
		\includegraphics[width=\columnwidth]{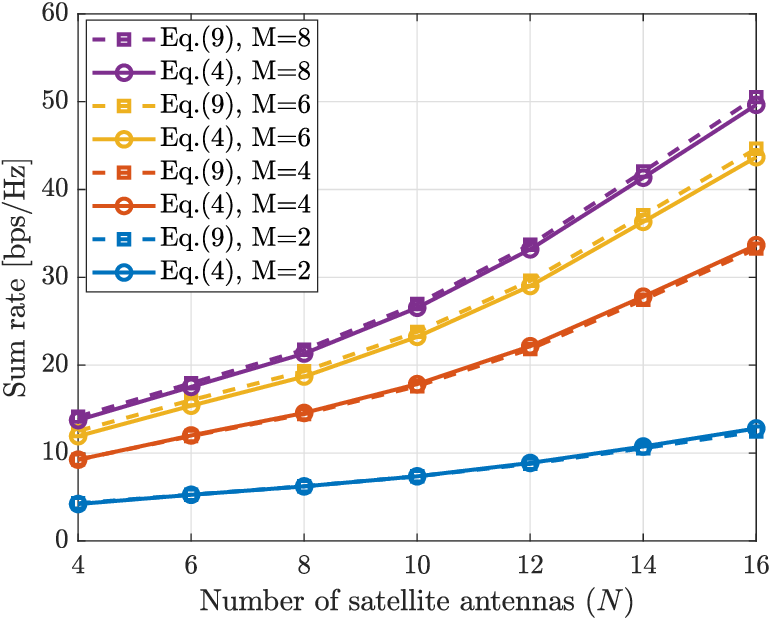}
		\caption{$L = 8$.}

	\end{subfigure}
	\caption{Evaluation of the approximate rate expression.}
\label{fig:approx_equality}
\vspace{-2mm}
\end{figure}

Fig.~\ref{fig:approx_equality} shows the sum rate evaluated via \eqref{eq:achievable_rate} and \eqref{eq:achievable_rate_approx} for different values of $N$, $M$, and $L$. We set $\vartheta_{\mathrm{s}} = 10^\circ$ and the sum rate is obtained by averaging over $10000$ Monte Carlo trials.
In this simulation, we employ MMSE precoding. 
The results show that \eqref{eq:achievable_rate_approx} provides an acceptable approximation for the actual achievable rate. Although the approximation is not always tight, \eqref{eq:achievable_rate} and \eqref{eq:achievable_rate_approx} show similar trends with varying parameters. This makes the approximated achievable rate a reasonable choice for precoder design and subsequent rate analysis.   

\begin{figure}
    \centering
    \includegraphics[width=0.67\linewidth]{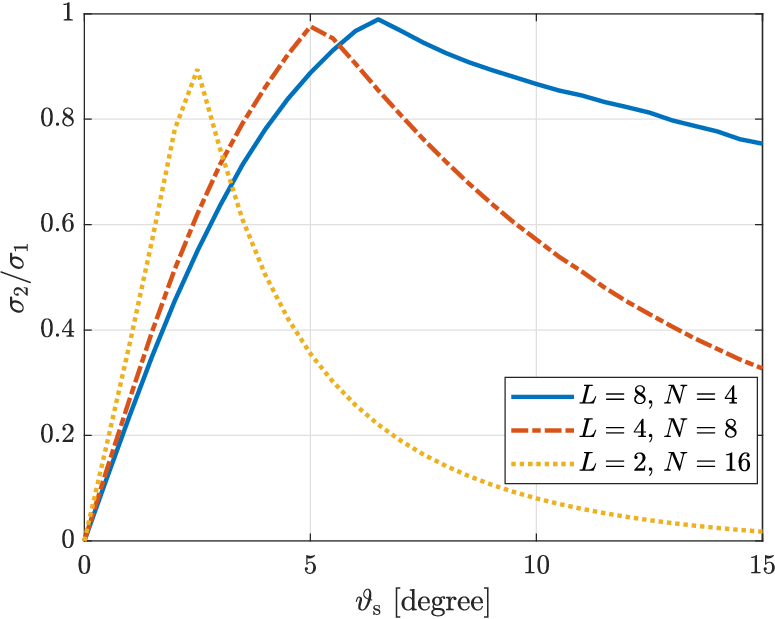}
    \caption{Effect of satellite angular spread on singular values. }
\label{fig:singular_values}
\vspace{-4mm}
\end{figure}

To illustrate the spatial multiplexing capability for the UEs, Fig.~\ref{fig:singular_values} plots the ratio $\sigma_2/\sigma_1$ versus $\vartheta_{\mathrm{s}}$ for different numbers of SATs, where $\sigma_1$ and $\sigma_2$ are the largest and second-largest singular values of the channel between a randomly chosen UE and all SATs. The total number of transmit antennas is fixed at $32$ for all cases. The point $\vartheta_{\mathrm{s}} = 0^\circ$ corresponds to all antennas being co-located.
The results show that increasing $\vartheta_{\mathrm{s}}$ initially improves the ratio $\sigma_2/\sigma_1$ due to reduced channel correlation and improved orthogonality between the channels from different SATs to the UE. 
After a certain point, the ratio starts to decrease. This is because beyond the optimal angular separation, the LoS geometry makes SAT-UE channels less distinct, which increases their correlation and degrades multiplexing. 
 We set   $\vartheta_s = 5^\circ$ in the following simulations.

\begin{figure}
    \centering
    \includegraphics[width=0.67\linewidth]{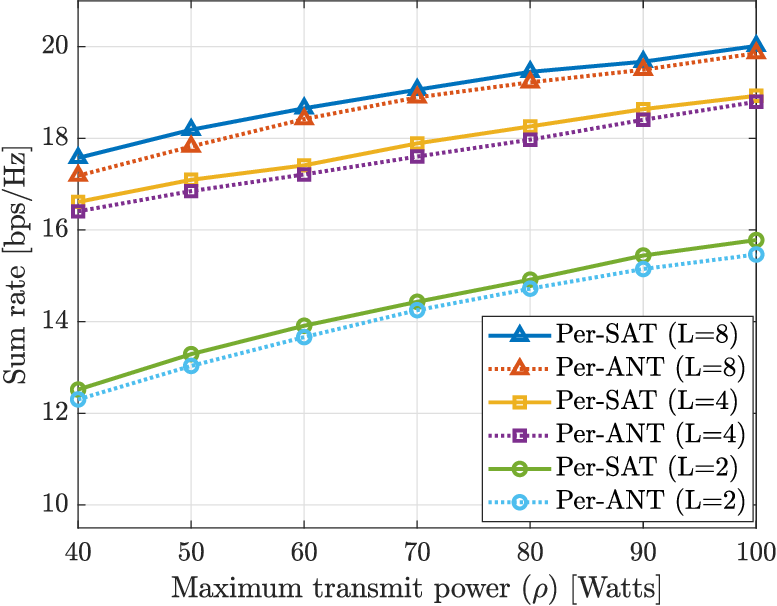}
    \caption{Sum rate versus power for varying numbers of SATs.}
\label{fig:rate_vs_power_L}
\vspace{-2mm}
\end{figure}
Fig.~\ref{fig:rate_vs_power_L} depicts the sum rate as a function of transmit power for both cases of per-SAT and per-antenna power constraints. In the latter case, the transmit power of each SAT is divided by 
$N$, and this value is taken as the budget for each antenna.
As expected, the average sum rate increases with both higher transmit power and a greater number of SATs, due to increased received power, enhanced beamforming gain, and improved interference suppression. We observe that the performance improvement from $L = 2$ to $L = 4$ is greater than the improvement from $L = 4$ to $L = 8$. This is because increasing the number of SATs from $2$ to $4$ significantly enhances beamforming gain and interference suppression.  While further doubling the number of SATs to $8$ continues to improve these aspects, the additional gains are smaller due to diminishing returns from spatial diversity, beamforming gain, and interference suppression capability. We can also see that the per-SAT power constraint case slightly outperforms the per-antenna case due to its greater power allocation flexibility.

\begin{figure}
    \centering
    \includegraphics[width=0.67\linewidth]{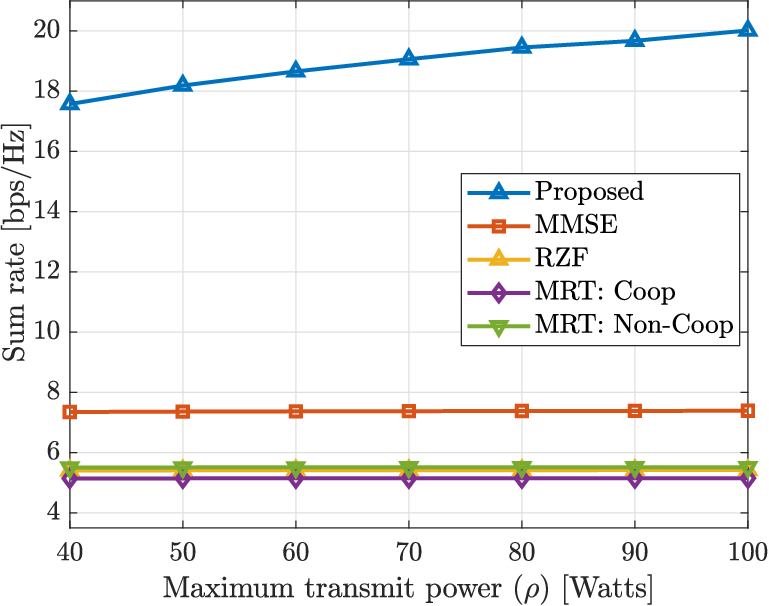}
    \caption{Sum rate versus power for different precoding designs.}
\label{fig:all_precoding}
\vspace{-4mm}
\end{figure}

Next, we compare the performance of the proposed scheme with traditional precoding designs, including MMSE, regularized zero-forcing (RZF), and maximum ratio transmission (MRT). We consider a per-SAT power constraint and $L = 8$ SATs, and add a non-cooperative MRT benchmark, where each SAT serves exactly one UE: starting with UE 1, we assign the closest available SAT as its serving SAT, remove that SAT from the pool, and continue sequentially through the UEs, each time choosing the closest remaining SAT. 
We can see that benchmarks yield little improvement with higher power due to their poor interference management. The proposed precoding substantially outperforms the benchmarks by iteratively updating precoders, receive combiners, and weights to better shape and cancel interference. Interestingly,  non-cooperative MRT outperforms cooperative MRT and RZF, as assigning each UE to a single SAT limits the extent of inter-user interference.  

\section{Concluding Remarks}
This paper investigated a multi-user distributed satellite communication with multi-antenna users. We optimized satellite precoding matrices under per-satellite and per-antenna power constraints using statistical channel information. The optimization employed an approximate rate function that, according to numerical results, is sufficiently accurate under realistic conditions.
Results indicated that multi-antenna users can support multiple data streams when satellite angular separation is properly selected. Moreover, our comparisons showed that the proposed precoding design significantly outperforms baseline methods due to superior interference suppression, which is critical in multi-user distributed environments.  

\bibliographystyle{IEEEtran}
\bibliography{refs} 

@IEEEtranBSTCTL{BSTcontrol,
  CTLuse_forced_etal       = "yes",
  CTLmax_names_forced_etal = "4",   
  CTLnames_show_etal       = "1"   
}

@ARTICLE{Zhang2025Enabling,
  author={Zhang, Yuchen and Al-Naffouri, Tareq Y.},
  journal={IEEE Transactions on Wireless Communications}, 
  title={Enabling Scalable Distributed Beamforming via Networked {LEO} Satellites Toward 6{G}}, 
  year={2026},
  volume={25},
  number={},
  pages={6666-6680},
  doi={10.1109/TWC.2025.3626203}}

@article{zhang2014power,
  title={Power scaling of uplink massive {MIMO} systems with arbitrary-rank channel means},
  author={Zhang, Qi and Jin, Shi and Wong, Kai-Kit and Zhu, Hongbo and Matthaiou, Michail},
  journal={IEEE Journal of Selected Topics in Signal Processing},
  volume={8},
  number={5},
  pages={966--981},
  year={2014},
  publisher={IEEE}
}

@ARTICLE{alsenwi2024robust,
  author={Alsenwi, Madyan and Lagunas, Eva and Chatzinotas, Symeon},
  journal={IEEE Transactions on Vehicular Technology}, 
  title={Robust Beamforming for Massive {MIMO LEO} Satellite Communications: A Risk-Aware Learning Framework}, 
  year={2024},
  volume={73},
  number={5},
  pages={6560-6571},
  doi={10.1109/TVT.2023.3338065}}

@article{demir2021foundations,
  title={Foundations of user-centric cell-free massive {MIMO}},
  author={Demir, {\"O}zlem Tugfe and Bj{\"o}rnson, Emil and Sanguinetti, Luca and others},
  journal={Foundations and Trends{\textregistered} in Signal Processing},
  volume={14},
  number={3-4},
  pages={162--472},
  year={2021},
  publisher={Now Publishers, Inc.}
}

@ARTICLE{Shi2011,
  author={Q. Shi and M. Razaviyayn and Z-Q. Luo and C. He},
  journal={IEEE Transactions on Signal Processing}, 
  title={An Iteratively Weighted {MMSE} Approach to Distributed Sum-Utility Maximization for a {MIMO} Interfering Broadcast Channel},
  year={2011},
  volume={59},
  number={9},
  pages={4331-4340}}

@article{wp5d2023m,
  title={M. 2160: framework and overall objectives of the future development of {IMT} for 2030 and beyond},
  author={ ITU-R},
month ={[Online]. Available: https://www.itu.int/rec/R-REC-M.2160/en},
  year={2023}
}

@book{grotschel2012geometric,
  title={Geometric algorithms and combinatorial optimization},
  author={Gr{\"o}tschel, Martin and Lov{\'a}sz, L{\'a}szl{\'o} and Schrijver, Alexander},
  volume={2},
  year={2012},
  publisher={Springer Science \& Business Media}
}

@inproceedings{humadi2024distributed,
  title={Distributed Massive {MIMO} System with Dynamic Clustering in {LEO} Satellite Networks},
  author={Humadi, Khaled and Kurt, Gunes Karabulut and Yanikomeroglu, Halim},
  booktitle={2024 6th International Conference on Communications, Signal Processing, and their Applications (ICCSPA)},
  pages={1--6},
  year={2024}
}

@inproceedings{ha2024user,
  title={User-centric beam selection and precoding design for coordinated multiple-satellite systems},
  author={Ha, Vu Nguyen and Nguyen, Duy HN and Duncan, Juan C-M and Gonzalez-Rios, Jorge L and Peralvo, Juan A V{\'a}squez and Eappen, Geoffrey and Garces-Socarras, Luis M and Palisetty, Rakesh and Chatzinotas, Symeon and Ottersten, Bj{\"o}rn},
  booktitle={2024 IEEE 35th International Symposium on Personal, Indoor and Mobile Radio Communications (PIMRC)},
  pages={1--6},
  year={2024},
}

@ARTICLE{wang2025multiple,
  author={Wang, Qi and Chen, Xiaoming and Qi, Qiao and Li, Mili and Gerstacker, Wolfgang},
  journal={IEEE Transactions on Wireless Communications}, 
  title={Multiple-Satellite Cooperative Information Communication and Location Sensing in {LEO} Satellite Constellations}, 
  year={2025},
  volume={24},
  number={4},
  pages={3346-3361},
  doi={10.1109/TWC.2025.3530083}}

@article{perez2019signal,
  title={Signal processing for high-throughput satellites: Challenges in new interference-limited scenarios},
  author={Perez-Neira, Ana I and Vazquez, Miguel Angel and Shankar, MR Bhavani and Maleki, Sina and Chatzinotas, Symeon},
  journal={IEEE Signal Processing Magazine},
  volume={36},
  number={4},
  pages={112--131},
  year={2019},
  publisher={IEEE}
}

@article{you2022hybrid,
  title={Hybrid analog/digital precoding for downlink massive {MIMO LEO} satellite communications},
  author={You, Li and Qiang, Xiaoyu and Li, Ke-Xin and Tsinos, Christos G and Wang, Wenjin and Gao, Xiqi and Ottersten, Bj{\"o}rn},
  journal={IEEE Transactions on Wireless Communications},
  volume={21},
  number={8},
  pages={5962--5976},
  year={2022},
  publisher={IEEE}
}

@article{Tseng2001,
	Author = {P. Tseng },
  journal   = {Journal of Optimization Theory and Applications},
title={Convergence of a Block Coordinate Descent
Method for Nondifferentiable Minimization}, 
year={2001}, 
volume={109}, 
number={3}, 
pages={475-494}}

@article{you2020massive,
  title={Massive {MIMO} transmission for {LEO} satellite communications},
  author={You, Li and Li, Ke-Xin and Wang, Jiaheng and Gao, Xiqi and Xia, Xiang-Gen and Ottersten, Bj{\"o}rn},
  journal={IEEE Journal on Selected Areas in Communications},
  volume={38},
  number={8},
  pages={1851--1865},
  year={2020},
  publisher={IEEE}
}
\end{document}